\documentstyle[12pt]{article}

\begin{document}

\renewcommand{\thefootnote}{\fnsymbol{footnote}}

\begin{titlepage}

\begin{flushright}
IJS-TP-95/11\\
NUHEP-TH-95-06\\
30 June  1995\\
\end{flushright}

\vspace{.5cm}

\begin{center}
{\Large \bf  Charm meson radiative decays\\}

\vspace{1.5cm}

{\large \bf B. Bajc $^{a}$, S. Fajfer $^{a}$ and
Robert J. Oakes $^{b}$\\}

\vspace{.5cm}

{\it a) Physics Department, Institute "J.Stefan", Jamova 39,
61111 Ljubljana, Slovenia\\}

\vspace{.5cm}

{\it b) Department of Physics and Astronomy,
Northwestern University, Evanston, Il 60208
U.S.A.}

\vspace{1cm}

\end{center}

\centerline{\large \bf ABSTRACT}

\vspace{0.5cm}

Combining heavy quark effective theory
and the chiral Lagrangian approach we investigate D meson radiative decays.
First, we reanalyse $D^{*} \rightarrow D \gamma$
decays using heavy quark spin symmetry, chiral symmetry
Lagrangian, but including also the light vector
mesons. Then, we calculate branching ratios $D \rightarrow V \gamma$.
We make some comments on the Cabbibo suppressed decays
$D \to \rho / \omega \gamma$ and the
branching ratios of
$D \rightarrow P P \gamma$ decays.

\end{titlepage}

\setlength {\baselineskip}{0.75truecm}
\parindent=3pt  % no indention of new paragraphs
\setcounter{footnote}{1}    % start footnotes at dagger instead
			     % of '*' (article style only)

\renewcommand{\thefootnote}{\arabic{footnote}}
\setcounter{footnote}{0}

\vspace{.5cm}

%\begin{center}
%{\bf 1.Introduction}
%\end{center}

Experimentally radiative decays of $D$ mesons
have not been measured, while the known branching ratios of
$D^{*}$ radiative decays \cite{CG,ABJ}
can be described using the
combination of
heavy quark effective theory and chiral Lagrangians
\cite{IW}-\cite{SCHECHTER}.
%\cite{IW,BD,YCC,LLY,PC,HG1,HG2,FGL,G1,G2,G3,BFO,SCHECHTER}.

The strong interaction meson Lagrangian for the light
pseudoscalar octet and heavy pseudoscalar
and vector triplets in the chiral and heavy
quark limits was first written down by Wise \cite{MW}
(see also \cite{BD}).
The electromagnetic
interactions between these
mesons was described in \cite{CG,ABJ,LLY}.
The octet of light vector mesons
was included in the Wise Lagrangian
\cite{MW} later by Casalbuoni, et al.
\cite{G2} as the gauge particles associated
with the hidden symmetry group SU(3)$_H$
\cite{BKY}.
The light pseudoscalar mesons are described by the
$3 \times 3$ unitary matrix $ u = exp (\frac{i \Pi}{f})$, ($ f \simeq 132$ MeV)
with $\Pi$ the usual pseudoscalar matrix
as in ref. \cite{BFO}, while
the octet of light vector mesons is described
by the $3 \times 3$ matrix
${\hat \rho}_\mu  =  i {g_V \over \sqrt{2}} \rho_\mu$
where $g_V$ ($\simeq 5.8 \sqrt{2/a}$ with $a=2$ in
the case of exact vector dominance) is the coupling
constant of the vector meson self--interaction
\cite{BKY} and $\rho_\mu$ the vector meson matrix.
The heavy mesons are $Q{\bar q}^{a}$ ground states,
where $Q$ is a c quark and
$q^{1} = u$, $q^{2} = d$ and $q^{3} = s$.
In the heavy quark limit they are
described by $4 \times 4$
matrix $H_{a}$ $(a = 1,2,3)$ \cite{MW}

\begin{eqnarray}
\label{esix}
H_a& = & \frac{1}{2} (1 + \!\!\not{\! v}) (P_{a\mu}^{*}
\gamma^{\mu} - P_{a} \gamma_{5})
\end{eqnarray}

where $P_{a\mu}^{*}$ and $P_{a}$ annihilate,
respectively, a spin-one and spin-zero meson
$Q{\bar q}^{a}$ of velocity $v_{\mu}$.
Following the procedure of \cite{BFO} we
write down the even parity
strong and electromagnetic
Lagrangian for heavy and light pseudoscalar and
vector mesons:

\begin{eqnarray}
\label{efifteen}
{\cal L}_{even} & = & {\cal L}_{light} -
{1 \over 4} (\partial_\mu B_\nu - \partial_\nu B_\mu)^2 +
i Tr (H_{a} v_{\mu} D^{\mu} {\bar H}_{a})\nonumber\\
& + &i g Tr [H_{b} \gamma_{\mu} \gamma_{5}
({\cal A}^{\mu})_{ba} {\bar H}_{a}]
 +  i \beta Tr [H_{b} v_{\mu} ({\cal V}^{\mu}
- {\hat \rho}_{\mu})_{ba} {\bar H}_{a}]\\
& + &  {\beta^2 \over 2 f^2 a}
Tr ({\bar H}_b H_a {\bar H}_a H_b)\nonumber
\end{eqnarray}

with

\begin{eqnarray}
\label{esixteen}
{\cal L}_{light} & = & -{f^2 \over 2}
\{tr({\cal A}_\mu {\cal A}^\mu) +
a\, tr[({\cal V}_\mu - {\hat \rho}_\mu)^2]\}\nonumber\\
& + & {1 \over 2 g_V^2} tr[F_{\mu \nu}({\hat \rho})
F^{\mu \nu}({\hat \rho})]
\end{eqnarray}

where $B_{\mu}$, ${\cal A}_{\mu}$, ${\cal V}_{\mu}$,  $D_{\mu}$ are
defined in ref. \cite{BFO}.
In equation (\ref{efifteen}) $g$ and $\beta$ are
constants which should be determined from
experimental data \cite{CG,ABJ,G1,G2,G3}.
The constant $a$ in (\ref{efifteen})-(\ref{esixteen})
is in principle a free parameter, but we shall
fix it by assuming exact
vector dominance \cite{BKY}, for which
$a=2$.
The electromagnetic field can couple to the
mesons also through the anomalous interaction;
i.e., through the odd parity Lagrangian.
We write down the two
contributions of the the odd parity Lagrangian which are significant for our
calculation \cite{WZ,EW}

\begin{eqnarray}
\label{enineteen}
{\cal L}^{(1)}_{odd} & = & -4 \frac{C_{VV\Pi}}{f} \epsilon
^{\mu \nu \alpha \beta}Tr (\partial_{\mu}
{\rho}_{\nu} \partial_{\alpha}{\rho}_{\beta} \Pi)
\end{eqnarray}

\begin{eqnarray}
\label{etwenty}
{\cal L}^{(2)}_{odd} & = & -4 e {\sqrt 2}
\frac{C_{V\pi\gamma}}{f} \epsilon
^{\mu \nu \alpha \beta}
Tr (\{ \partial_{\mu}{\rho}_{\nu},
\Pi \} Q \partial_{\alpha} B_{\beta})
\end{eqnarray}

Equation (\ref{etwenty}) and equation
(\ref{enineteen}), together with vector dominance
couplings

\begin{eqnarray}
\label{etwentyone}
{\cal L}_{V-\gamma} & = & - m_V^2 {e \over g_V} B_{\mu}
(\rho^{0\mu} + {1 \over 3} \omega^{\mu} -
{\sqrt{2} \over 3} \Phi^{\mu})
\end{eqnarray}

which come from the second term in (\ref{esixteen}),
describe the anomalous type electromagnetic
interactions in the light sector.
The interactions of light
vector mesons (and of photons via vector-dominance
(\ref{etwentyone})), heavy pseudoscalars or
heavy vector $D$ mesons
is introduced through
the higher dimensional
invariant operator
\begin{eqnarray}
\label{etwentytwo}
{\cal L}_{1} & = & i {\lambda} Tr [H_{a}\sigma_{\mu \nu}
F^{\mu \nu} (\hat \rho)_{ab} {\bar H_{b}}]
\end{eqnarray}
The heavy quark-photon
interaction is genereted by the term

\begin{eqnarray}
\label{etwentythree}
{\cal L}_{2} & = & - \lambda^{\prime}
e Tr [H_{a}\sigma_{\mu \nu}
F^{\mu \nu} (B) {\bar H_{a}}]
\end{eqnarray}

According to quark models
the parameter $|\lambda^{\prime}|$ can be
approximately related to
the charm quark magnetic moment via
$1/(6 m_c)$ \cite{CG,ABJ,YCC,LLY}.
The experimentally measured  \cite{PD,expr}  branching fractions are
$R_{\gamma}^0 = \Gamma (D^{*0} \to D^0 \gamma) /
\Gamma (D^{*0} \to D^0 \pi^0)=0.572 \pm 0.057 \pm 0.081$.
and $R_{\gamma}^+ = \Gamma (D^{*+} \to D^+ \gamma) /
\Gamma (D^{*+} \to D^+ \pi^0)=0.035 \pm 0.047 \pm 0.052$.
With our Lagrangians  these branching ratios are calculated to be

\begin{eqnarray}
\label{etwentyfour}
R_{\gamma}^0 & = & 64 \pi f^2 \alpha_{EM}
({\lambda' \over g} + {2 \over 3} {\lambda \over g} )^2
({p_{\gamma}^0 \over p_{\pi}^0})^3
\end{eqnarray}

\begin{eqnarray}
\label{etwentyfive}
R_{\gamma}^+ & = & 64 \pi f^2 \alpha_{EM}
({\lambda' \over g} - {1 \over 3} {\lambda \over g} )^2
({p_{\gamma}^+ \over p_{\pi}^+})^3
\end{eqnarray}

{}From the experimental data, we have found \cite{BFO}
\begin{eqnarray}
\label{etwentyeight}
|{\lambda' \over g}+{2 \over 3}{\lambda \over g}| & = &
(0.863 \pm 0.075) \hbox{GeV}^{-1}
\end{eqnarray}

and

\begin{eqnarray}
\label{etwentynine}
|{\lambda' \over g}-{1 \over 3}{\lambda \over g}| & = &
(0.089 \pm 0.178) \hbox{GeV}^{-1}
\end{eqnarray}

In our approach both $\lambda$ and $\lambda^{\prime}$ are considered as
purely phenomenological.

The part of the weak Lagrangian for the pseudoscalar and
vector, light and heavy mesons, which we will use, can be
written as \cite{KXC,WSB1}

\begin{eqnarray}
\label{ethirtyfive}
{\cal L}_{W}^{eff}(\Delta c = \Delta s = 1) =
-{G \over \sqrt{2}} V_{ud} V_{cs}^{*}
[ & a_{1} ({\bar u}d)_{V-A}^\mu ({\bar s}c )_{V-A,\mu} + &
\nonumber\\
& a_{2} ({\bar s}d)_{V-A}^\mu ({\bar u}c )_{V-A,\mu} & ]
\end{eqnarray}

where $V_{ud}$, etc. are the relevant $CKM$
mixing paremeters, while $a_{1}$ and $a_{2}$ are the
QCD Wilson coefficients, which depend on a scale
$\mu$. One expects the scale to be the heavy quark mass
and we take $\mu \simeq 1.5$ GeV which
gives $a_{1} = 1.2$ and  $a_{2} = -0.5$,
with an approximate  $20\%$ error.
Many heavy meson weak nonleptonic
amplitudes \cite{KXC,WSB1,G4} have been
calculated using the factorization approximation.
The quark currents are approximated by the
corresponding meson currents \cite{buras} defined
later in eqs. (\ref{ethirty}) and (\ref{ethirtyfour}):

\begin{eqnarray}
\label{ethirtysix}
({\bar q}_a Q)_{V-A}^\mu & \equiv &
{\bar q}_a \gamma^\mu (1-\gamma^5) Q \simeq
{J_Q}_{a}^\mu
\end{eqnarray}

\begin{eqnarray}
\label{ethirtyseven}
({\bar q}_b q_a)_{V-A}^\mu & \equiv &
{\bar q}_b \gamma^\mu (1-\gamma^5) q_a \simeq
{J_q}_{a b}^\mu
\end{eqnarray}

In order to describe weak currents we use the
"bosonised" versions like \cite{BFO,G2,MW}.
For the quark current with one heavy quark we will use
\begin{eqnarray}
\label{ethirty}
{J_Q}_{a}^{\mu} & = & \frac{1}{2} i f_{H} \sqrt{m_{H}} Tr [\gamma^{\mu}
(1 - \gamma_{5})H_{b}u_{ba}^{\dag}]
\end{eqnarray}
The light meson part of the weak current is

\begin{eqnarray}
\label{ethirtyfour}
{J_q}^{\mu} & = & i f^2 u [{\cal A}^{\mu}+
a ( {\cal V}^{\mu}-{\hat \rho}^{\mu}) ] u^{\dag}
\end{eqnarray}

The simplest radiative decays of D mesons are
into a light meson and a photon. Since the
process $D \to P \gamma$ (P is a light pseudoscalar)
is forbidden due to the requirement of gauge
invariance and chiral symmetry \cite{Ecker},
as well as angular momentum conservation,
we will concentrate on the $D\rightarrow
V \gamma$ (V is a light vector meson) decays.
We consider the only two processes which are
possible at tree level and are not Cabibbo supressed,
namely $D^{0}\rightarrow \bar{K}^{*0} \gamma$ and
$D^{s+}\rightarrow \rho^+ \gamma$. Both processes
have contributions from the odd-parity interaction
Lagrangian. The second one has, in addition, a direct
emission term, due to
the charged initial and final mesons.

In our numerical calculations we used the following
numerical values $C_{VV\Pi} = 0.423$,
$C_{V\Pi\gamma} = -3.26\times 10^{-2}$,
\cite{BOS,FSO}, $g_{V} = 5.8$ \cite{G2},
$f \simeq f_{\pi} = 132$ MeV, and
the other decay constants $f_{P,V}$
were taken from \cite{G4}.
It is straightforward to calculate the decay widths.
The result, of course, depends on which
numerical value we take for $(\lambda' +
2 \lambda /3)$ and $(\lambda'-\lambda/3)$.

Computing these decay widths using $\lambda$ and $\lambda'$
we have to point out that the
$\frac{1}{m_{c}}$ corrections, coming from light-quark current,
effectively included into the $\lambda'$ parameter, are not
necesserily the same as in the case
$D^{*} \rightarrow D \gamma$ decay. Of course, this
uncertainty unfortunately
increases the theoretical and experimental uncertainties
already present
in the calculation of the $D \rightarrow V \gamma$.

On figs. 1 and 2
the decay widths for $D^0\to\bar{K}^{*0}\gamma$ and
$D^{s+}\to\rho^+\gamma$ are shown as functions of
the combinations $\lambda'+2\lambda/3$ and
$\lambda'-\lambda/3$ respectively. The full (dashed) lines
denotes the values for these combinations, which
are allowed (forbidden) by experimental constraints $(11)-(12)$
together with $|g|=0.57\pm 0.13$ \cite{G3}.

An interesting feature can be seen from fig. 1:
a not very precise
measurement of the $D^0 \to \bar{K}^{*0} \gamma$
decay width is sufficient to differentiate between
positive or negative solutions for $\lambda'+2\lambda/3$,
which are predicted to be of one order of magnitude different.
Given the total $D^0$ decay width
$\Gamma_{TOT}(D^0)\approx 0.0016$ eV
\cite{PD} the branching ratio
is constrained by $2\times 10^{-4}<B<4\times 10^{-4}$ for
positive and $B<0.4\times 10^{-4}$ for negative values
of $\lambda'+2\lambda/3$. Our prediction is consistent
with the nonrelativistic quark model \cite{AK} and with the
order of magnitude estimate of ref. \cite{CHENG}, but not
with the analysis of Burdman et al. \cite{BURDMAN}, who
get a branching ratio somewhere between our allowed regions.

A less dramatic situation is obtained for the
decay $D^{s+}\to\rho^+\gamma$ in fig. 2,
where a not precise determination of the partial
width is not enough to further constrain the
combination $\lambda'-\lambda/3$. From
$\Gamma_{TOT}(D^{s+})\approx 0.0014$ eV \cite{PD}
we see that the branching ratio for this decay
is in the range $2\times 10^{-4}<B<7\times 10^{-4}$.
The result is similar to ref. \cite{BURDMAN}, but it
is one order of magnitude larger than the one in \cite{AK}.

In ref. \cite{BIGI}, \cite{BIGI1}
it was noticed that a nice bonus
can be obtained in measuring the charm meson decay
$D^0\to\rho^0(\omega)\gamma$, which can get
contributions from New Physics by $c\to u\gamma$
transitions, while the same contributions
are absent in the $D^0\to\bar{K}^{*0}
\gamma$ decay. The authors of \cite{BIGI1} claim that a
discrepancy between the experimental measured ratio
of the two decays and the
theoretical expectation, i.e. the violation of

\begin{equation}
\label{ratio}
R_{\rho(\omega)}={B(D\to\rho(\omega)\gamma)\over
B(D\to K^*\gamma)}={\tan{\theta_c}^2\over 2}
\end{equation}

\noindent
would be a clear sign of non-Standard Model physics.
Equation (\ref{ratio}), in which we have included a
factor of $1/2$ that was overlooked in refs. \cite{BIGI},
\cite{BIGI1}, can be derived from eq. (\ref{ethirtyfive})
in the factorization approximation.

In our approach, which we believe
to describe effectively the low-energy
physics of the Standard Model, relation
(\ref{ratio}) is correct in two cases:

\noindent
1) in the U(3) limit, where the masses of the
light pseudoscalar nonet are equal, as well as the masses
and the decay constants of the light vector nonet;

\noindent
2) in the case that the diagram with
the photon emission by the heavy meson
dominates, i.e. for positive values of
the experimentally allowed value for the combination
($\lambda'+2\lambda/3$) (see eq. (\ref{etwentyeight})),
providing that the vector decay constants are
approximately equal.

In terms of quark diagrams, there
are also penguin contributions, for example, to
$D\to\rho/\omega\gamma$, but not to $D\to K^*\gamma$.
In the following we will neglect them, since they were
found to be rather small \cite{BURDMAN}.

Our conclusion is, that the eq. (\ref{ratio})
may not be satisfied even in
the Standard Model, as is shown on
fig. 3: if the combination ($\lambda'+2\lambda/3$) turns out
to be negative, the ratio (\ref{ratio}) can be anything between
$0$ and $\infty$.
As shown in the case $D\to K^*\gamma$, the negative values
of ($\lambda'+2\lambda/3$) can cause a destructive interference
between the photon emission by the heavy and the light meson.
A similar effect is possible also in the decay
$D\to\rho(\omega)\gamma$, only that the zero can be achieved
at a different value of ($\lambda'+2\lambda/3$), because of the
U(3) breaking. It is clear that such a situation
does not allow us to conclude anything about some New Physics.
Actually the situation would be in this case even worse, because
negative values of ($\lambda'+2\lambda/3$) would mean a very
small branching ratio (see fig. 1) and so an
extremely difficult measurement.

If, on the other side, ($\lambda'+2\lambda/3$) turns out
to be positive, the decays are much easier to detect
experimentally, and also the theoretical situation is clearer,
since the curve is approaching the ideal theoretical value $1$
(fig. 3). A large disagreement with the theoretical
prediction (\ref{ratio}) would give in this case some sign
of New Physics. But even here one should take care, since
the amplitudes are approximately proportional to the decay
constants of the final vector meson. This can be seen, if
we calculate the decay $D^0\to\omega\gamma$ with the values
of the light vector decay constants taken from \cite{G4}:
$f_{K*}=f_\rho=221$ MeV and $f_\omega=156$ MeV. In this case
we gat a similar curve as in fig. 3 for
$D^0\to\rho^0\gamma$, but for large
positive values of ($\lambda'+2\lambda/3$) the ratio is
approaching a value of approximately $0.5$ instead of $1$.
The fact can be explained by the difference in the decay
constants, i.e. $(f_\omega/f_{K^*})^2\approx 0.5$.

The radiative decays $D\rightarrow P P \gamma$ have not been measured yet.
In ref. \cite{BIGI} it was pointed out that branching ratios of
D mesons are determined with $5-10 \%$ accuracy.
We analyse the branching ratios of $D\rightarrow P P \gamma$ decays in
order to see if they are measurable. As an example
we investigate the decay  $D\rightarrow K^{-} \pi^{+} \gamma$.
In this decay there are charged particles in the final state and therfore
we expect that the dominant contribution will come out from the
"bremsstrahlung".
We simply take the amplitude $|{\cal A}(D\rightarrow K^{-} \pi^{+})|$ from the
corresponding decay and write

\begin{eqnarray}
{\cal A}(D(P)\rightarrow K^{-}(k) \pi^{+}(p) \gamma (q) ) & = & - e {\cal A}
(D\rightarrow K^{-} \pi^{+})( \frac{\epsilon \cdot k}{k\cdot q}
- \frac{\epsilon \cdot p}{p\cdot q})
\label{twop1}
\end{eqnarray}

The integration over the phase space is done making
the cut in the pho\-ton
energy $\omega_{min}$. The re\-sult is pre\-sen\-ted  on fig. 4.
For the photon energy cut $\omega_{min}= 10$ MeV we
find  $B(D\rightarrow K^{-}\pi^{+} \gamma) =
1.6 \times 10^{-3}$, higher than the calculated
$B(D^{0}\rightarrow {\bar K}^{*0} \gamma) $.
Contrary to the bremsstrahlung in
$D^{0}\rightarrow K^{-} \pi^{+}\gamma$,
there are  only neutral pseudoscalar mesons in
the final state of the decay
$D\rightarrow {\bar K}^{0} \pi^{0} \gamma$,
and therefore the decay amplitude is
the so-called direct emission \cite{eck1}.
Usually the branching ratio is much smaller in
decays with direct emission than in decays where
the bremsstrahlung dominates \cite{eck1}.

We can conclude that our framework - the combination of
heavy quark symmetry and chiral symmetry - builds the effective
strong, weak and electromagnetic Lagrangian. Within this scheme the
calculated
$D\to V \gamma$ decay widths provide some guidence.
One should be careful in experimental searches,
since the bremsstrahlung
for $D^{0}\rightarrow K^{-} \pi^{+} \gamma$
leads to higher branching ratio than
for $D^{0}\rightarrow {\bar K}^{*0} \gamma$.
The hope for new physics in $D\to\rho/\omega\gamma$
coming from $c\to u \gamma$ transitions is lost since
the result obtained using  our approach might screen
possible signals coming from
nonminimal SUSY.

\vskip 0.5cm
{\it Acknowledgement.} This work was supported in part by the
Ministry of Science and Technology of the Republic
of Slovenia (B.B. and S.F.)
and by the U.S. Department
of Energy, Division of High Energy Physics,
under grant No. DE-FG02-91-ER4086 (R.J.O.).

\newpage
\centerline{FIGURES}

\vskip 1cm
\noindent
%Fig. 1: The experimentally allowed values for
%$\lambda/g$ and $\lambda'/g$ (shaded regions) are given
%by the conditions from the measured branching fractions
%$R_\gamma^0$ (full line) and $R_\gamma^+$ (dashed line).

\vskip 0.5cm
\noindent
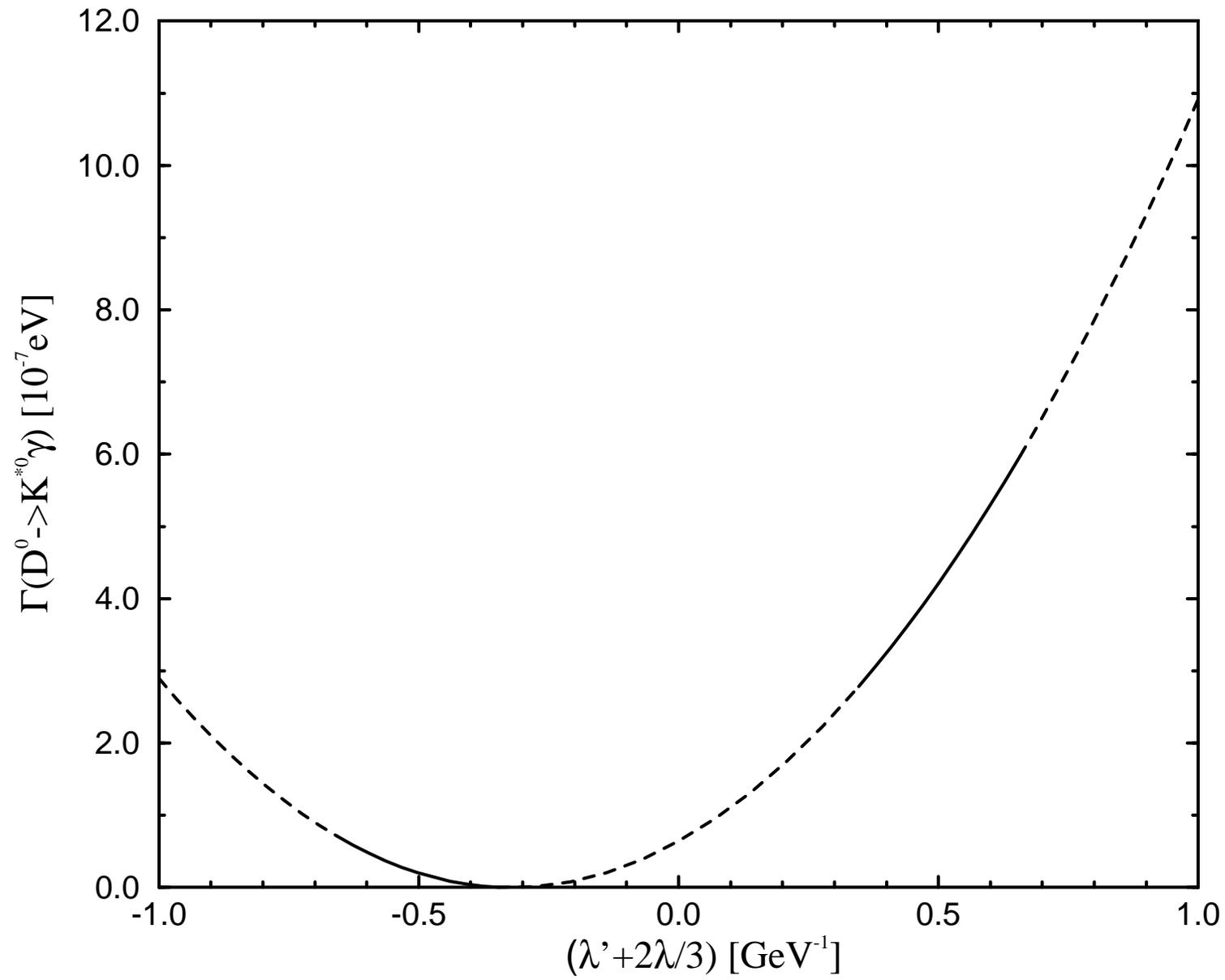
Fig. 1: The decay width for $D^0\to\bar{K}^{*0}\gamma$
as function of the combination $\lambda'+2\lambda/3$. The full
(dashed) lines denote the experimentally allowed (forbidden)
values for this combination.

\vskip 0.5cm
\noindent
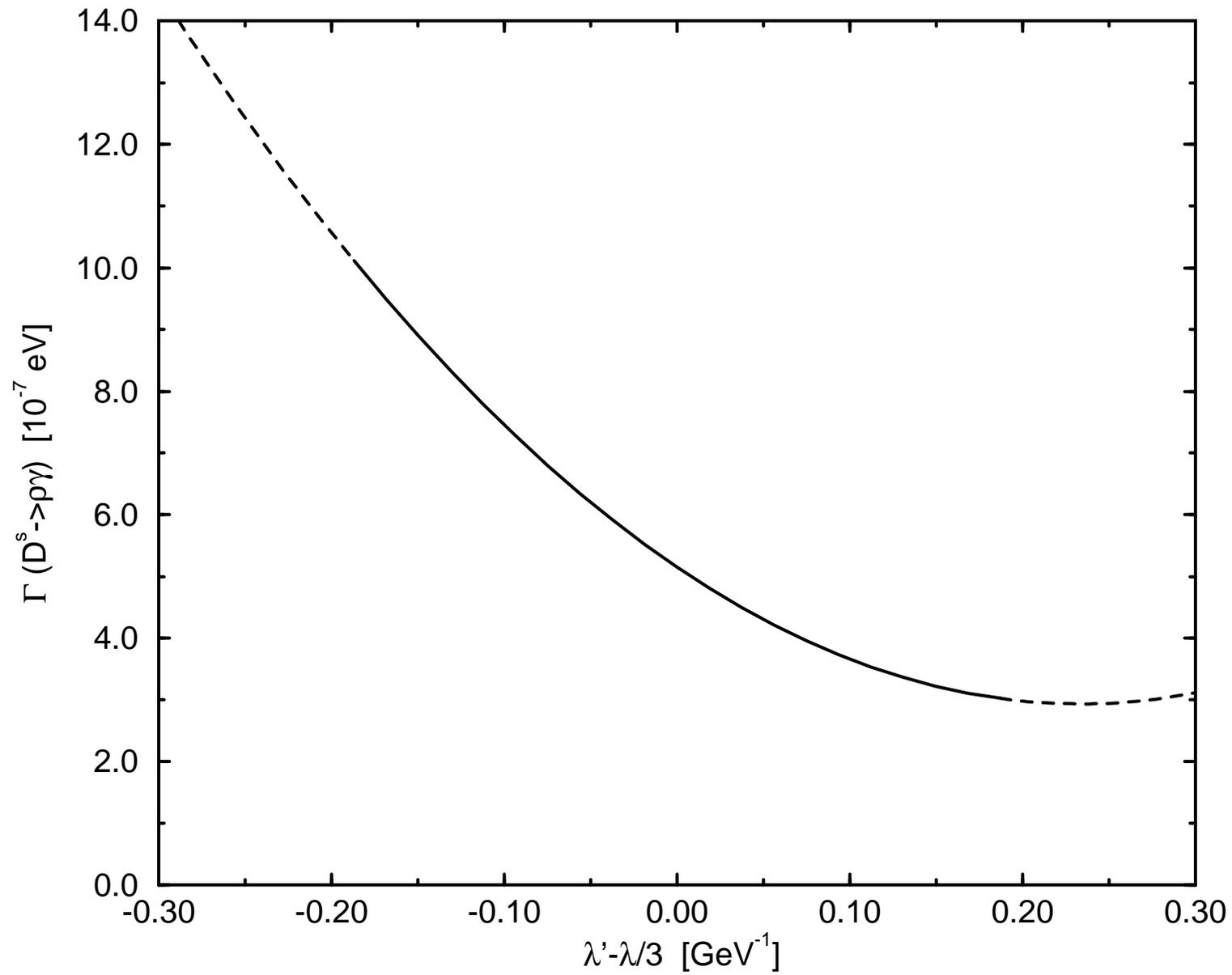
Fig. 2: The decay width for $D^{s}\to\rho\gamma$
as function of the combination $\lambda'-\lambda/3$.
The full
(dashed) line denotes the experimentally allowed (forbidden)
values for this combination.

\vskip 0.5cm
\noindent
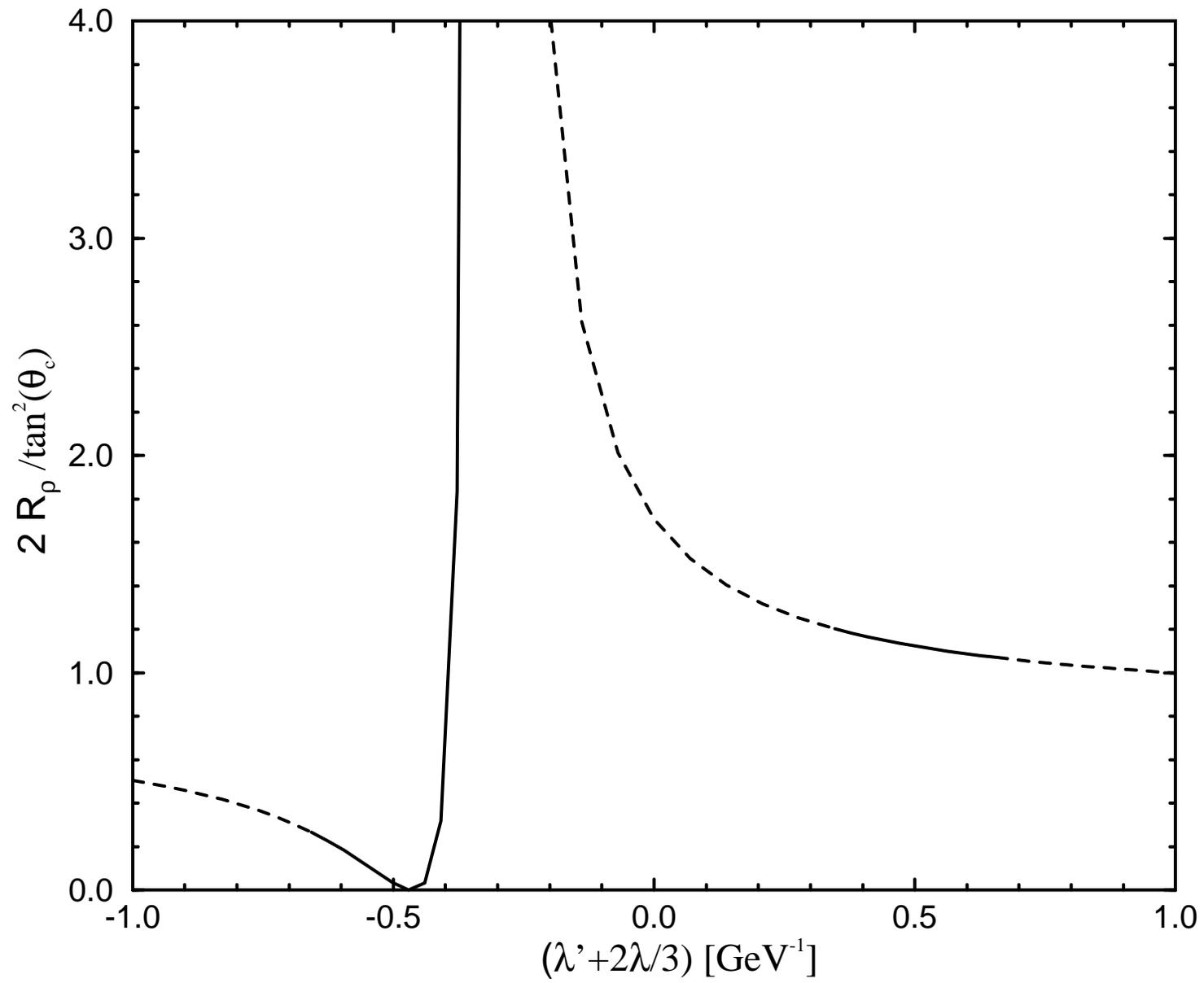
Fig. 3: The ratio of the decay rates for $D\to\rho\gamma$
and $D\to K^*\gamma$ times a constant factor $2/\tan{\theta_c}^2$
as function of the combination ($\lambda'+2\lambda/3$).
The full
(dashed) line denotes the experimentally allowed (forbidden)
values for this combination.
In the U(3) limit of the Standard Model
such a ratio should be equal to $1$.

\vskip 0.5cm
\noindent
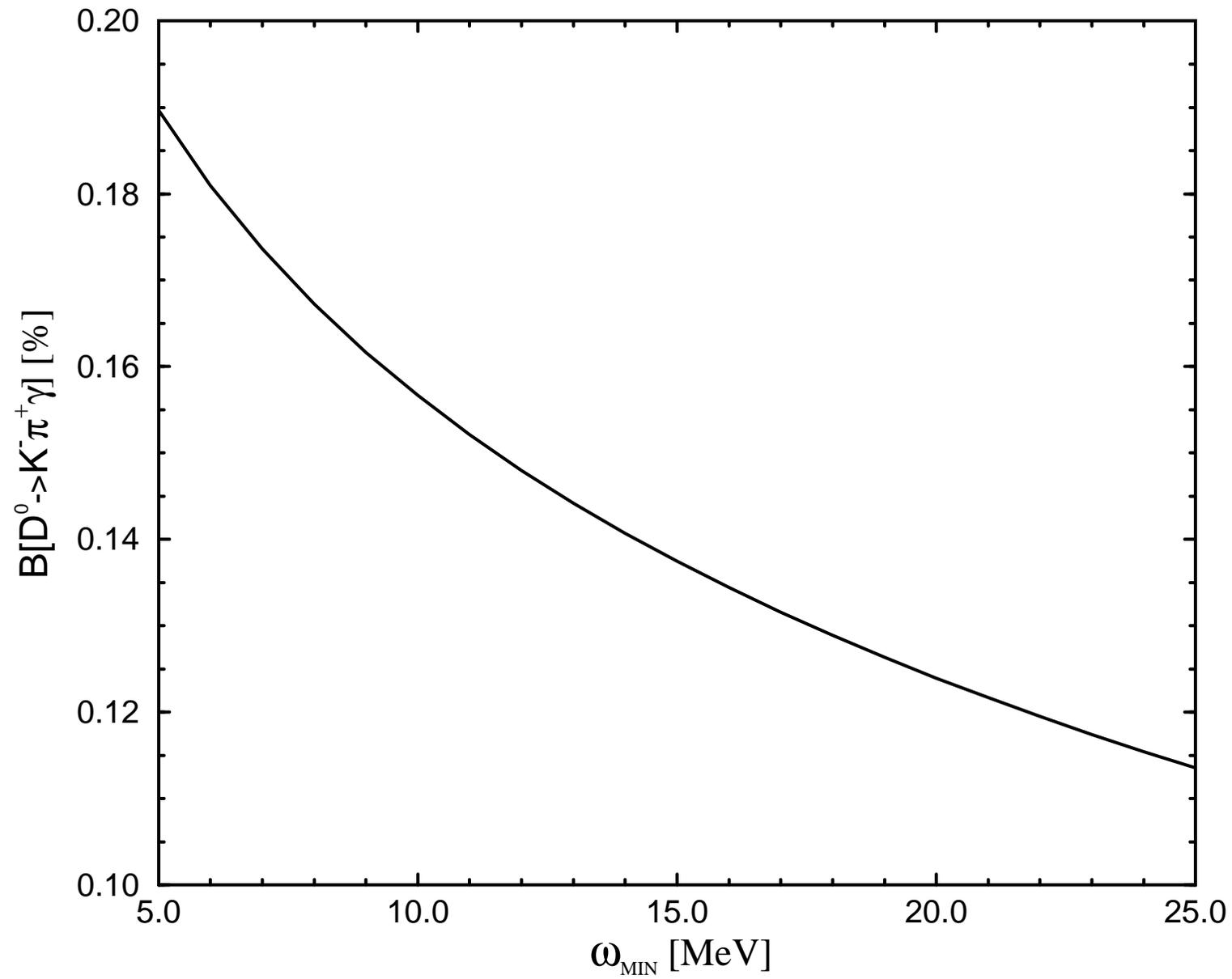
Fig. 4: The branching ratio of $D^{0} \to \pi^{+} K^{-} \gamma$
as function of the photon energy cut.


\vskip 0.5cm
\begin{thebibliography} {99}
\bibitem{CG} P. Cho and H. Georgi,
Phys. Lett. {\bf B296} (1992) 408.
\bibitem{ABJ} J. Amundson, C.G. Boyd,
E. Jenkins, M. Luke, A. Manohar,
J. Rosner, M. Savage and M. Wise,
Phys. Lett. {\bf B296} (1992) 415.
\bibitem{IW} N. Isgur and M. Wise,
Phys. Lett.{\bf B232} (1989) 113;
{\bf B237} (1990) 527.
\bibitem{BD} G. Burdman and J. Donoghue,
Phys. Lett.{\bf B280} (1992) 287;
\bibitem{YCC} T.M. Yan, H.Y. Cheng,
C.Y. Cheung, G.L. Lin, Y.C. Lin, H.L. Yu,
Phys. Rev. {\bf D46} (1992) 1148.
\bibitem{LLY}  H.Y. Cheng, C.Y. Cheung,
G.L. Lin, Y.C. Lin, T.M. Yan and H.L. Yu, Phys. Rev.
{\bf D47} (1993) 1030; {\bf D49} (1994) 2490.
\bibitem{PC} P. Cho, Nucl. Phys. {\bf B396} (1993) 183.
\bibitem{HG1} H. Georgi, Nucl. Phys. {\bf B348} (1991) 293.
\bibitem{HG2} H. Georgi, Phys. Lett. {\bf B240} (1990) 447.
\bibitem{FGL} A. Falk, B. Grinstein and M.Luke, Nucl. Phys.
{\bf B357} (1991) 185.
\bibitem{G1} R. Casalbuoni, A. Deandra, N.Di Bartolomeo, R. Gatto,
F. Feruglio, G. Nardulli, Phys. Lett. {\bf B294} (1992) 106.
\bibitem{G2} R. Casalbuoni, A. Deandra, N.Di. Bartolomeo, R. Gatto,
F. Feruglio, G. Nardulli, Phys. Lett. {\bf B292} (1992) 371.
\bibitem{G3} R. Casalbuoni, A. Deandra, N.Di Bartolomeo, R. Gatto,
F. Feruglio, G. Nardulli, Phys. Lett. {\bf B299} (1993) 139.
\bibitem{BFO} B. Bajc, S. Fajfer and R. J. Oakes,
Phys. Rev {\bf D51} (1995) 2230.
\bibitem{SCHECHTER} P. Jain, A. Momen, J. Schechter,
hep-ph/9406338.
\bibitem{MW} M. Wise, Phys. Rev. {\bf D45} (1992) 2188.
\bibitem{BKY} M. Bando, T. Kugo, S. Uehara,
K. Yamawaki and T. Yanagida, Phys. Rev. Lett.
{\bf 54} (1985) 1215;
M. Bando, T. Kugo, and K. Yamawaki,
Nucl. Phys {\bf B259} (1985) 493;
Phys. Rep. 164 (1988) 217.
\bibitem{WZ}  J. Wess and B. Zumino,
Phys. Lett. {\bf B37} (1971) 95.
\bibitem{EW} E. Witten, Nucl. Phys. {\bf223} (1983) 422.
\bibitem{PD} Review of Particle Properties 1994,
Phys. Rev. {\bf D 50} (1994) 1173.
\bibitem{expr} CLEO Collab., F.Butler at al.,
Phys. Rev. Lett. {\bf 69} (1992) 2041.
\bibitem{KXC} A.N. Kamal, Q.P. Xu and A.Czarnecki,
Phys. Rev. {\bf D49}(1994) 1330.
\bibitem{WSB1} M.Bauer, B. Stech and M. Wirbel,
Z. Phys. {\bf C34} (1987)103.
\bibitem{G4} A. Deandra, N. Di Bartolomeo,
R. Gatto and G. Nardulli,Phys. Lett. {\bf B318} (1993) 549.
\bibitem{buras} W.A. Bardeen, A.J. Buras and J.-M. G\'{e}rard,
Phys. Lett. {\bf B192} (1987) 138.
%\bibitem{BKY2} M. Bando, T. Kugo and K. Yamawaki,
%Prog. Theor. Phys. {\bf 73} (1985) 1541.
%\bibitem{LC} L.L.Chau, H.Y. Cheng,
%preprint ITP-B-93-49 and UCD-93-31,(1994).
\bibitem{Ecker} G. Ecker, A. Pich and E. de Rafael,
Nucl. Phys. {\bf B291}(1987) 692.
\bibitem{BOS} E. Braaten, R.J. Oakes and Sze-Man Tse,
Int. Jour. Mod. Phys. {\bf A5} (1990) 2737.
\bibitem{FSO} S. Fajfer, K. Suruliz and R.J. Oakes,
Phys. Rev. {\bf D46}(1992) 1195.
\bibitem{AK} P. Asthana and A.N. Kamal,
Phys. Rev. {\bf D43}(1991) 278.
\bibitem{CHENG} H.Y. Cheng, C.Y. Cheung, G.L. Lin, Y.C. Lin,
T.M. Yan, H.L. Yu, Phys. Rev. {\bf D51}(1995) 1199.
\bibitem{BURDMAN} G. Burdman, E. Golowich, J.L. Hewett,
S. Pakvasa, FERMILAB-Pub-94/412-T, UMHEP-415, SLAC-PUB-6692,
UH-511-811-94, hep-ph/9502329.
\bibitem{BIGI} I. I. Bigi, preprint CERN-TH.7370/94,
UND-HEP-94-BIG08.
%\bibitem{TP} T. Podobnik (ARGUS collaboration), private communication.
%\bibitem{bfo} B. Bajc, S. Fajfer and R.J. Oakes, in preparation.
\bibitem{BIGI1} I. Bigi, F. Gabbiani, A. Masiero, Z. Phys. {\bf C48}
(1990) 633.
\bibitem{eck1} G. D. Ambrosio et al., preprint
INFNNA-IV-94/24, DSF-94/24,
UWThPh-1994-20, Oct. 1994 and references therein.
\end{thebibliography}
\end{document}